# An Empirical Analysis on Remittances and Financial Development in Latin American Countries.

Sumaiya Binta Islam and Laboni Mondal

**Abstract**

Remittances became one of the driving forces of development for the countries all over the world, especially in the lower middle-income nations. This paper empirically investigates the association between remittance flows and financial development in 4 lower middle-income countries of Latin America. By using a panel data set from 1996–2019, the study revealed that remittances and financial development are positively associated in these countries. The study also discovered that foreign direct investment and inflation were positively correlated with financial development while trade openness had a negative association with financial development. Therefore, policymakers of these countries should implement and formulate such policies so that migrant workers would have the incentives to send money through formal channels, which will augment the effect of remittances on the recipient country.

## 1. Introduction

*1.1 Introduction*

The effects of Remittance flow in recipient countries have caught a great deal of attention in recent years. As remittance plays an important role in the economy by increasing the GDP volume, developing or underdeveloped countries considerably rely on it for their economic and financial development. For the circumstance of Latin America, remittance is a key component for their economic growth as they are the leading percipient of it. According to the Center for Latin American Monetary Studies (CEMLA), in 2017 Latin America and the Caribbean got $77.02 billion foreign remittances from their migrant workers. In developing countries, remittance flows, export earnings, foreign aid, foreign direct investment, financial inclusion etc. are important economic variables as these variables contribute to the development of a country (Biswas, 2023a; Das & Sethi, 2020; Fosu, 1990; Islam & Biswas, 2023).

There are 21 countries in Latin America, but for this study we have chosen 4 lower-income countries of this region, and the countries are Bolivia, Honduras, El Salvador, and Nicaragua. Among these countries, El Salvador is the highest remittance recipient country followed by Honduras and while Bolivia is the lowest recipient of remittances. Moreover, regarding the growth rate of remittances, Bolivia has a declining rate whereas other three countries have upward trends. Remittance is acknowledged to act as rescuer from greater economic shocks in times of the regional slowdown.

A wide range of studies have indicated and exhibited that remittances influence recipient countries in a variety of dimensions. Remittances helps in reducing poverty and inequality and contribute to economic growth of the recipient country. Because of remittances, individual household's living conditions change and holistically their growth is visible with upgraded financial solvency. Remittances received from migrant workers also benefit households to overcome credit constraints. Moreover, remittances also aid in boosting the nation's growth through foreign capital accumulation. Remittances intervene financial development in various aspects, a few of them are – i) it gives more access to capital markets with tools like diaspora bonds, ii) more access to financial services to household and businesses with checking and savings accounts, microfinance etc., and iii) facilitating transactions by enhancing the payment system with mobile money, Fintech, and risk mitigation. Hence, this paper analyzes the relationship between remittances and financial development in 4 countries of Latin America.

*1.2 Financial Development*

Before proceeding to the main paper, we would like to discuss the definition of financial development. Usually market, various banks and financial institutions, legal and regulatory framework altogether make the financial sector, which enable transactions and extend money to the economy. Accessibility regarding information of money deposit, withdrawal, taking loans, buying bonds or stocks from banks and intermediaries indicates the development and progress of this sector. Primarily the lower amount of cost incurring tells us the improvement of financial system. There are various measures of financial development, such as outstanding credit to the private sector relative to GDP, deposits to the private sector relative to GDP, broad money to GDP etc. In this study, we will use the broad money to GDP as the proxy of financial development.

## 2. Review of Literatures

There are a great number of theoretical and empirical papers that explored the effect of remittances on financial and economic development. The following section reviews some of the existing literatures in this regard.

Mundaca (2009) developed a theoretical model to analyze the effect of remittances on economic growth. By using a panel data set, the author tested the model and found that remittances had a significant positive effect on growth. But the effect of remittances became stronger when financial market development and remittances together entered growth equation.

Aggarwal et al. (2011) investigate the relationship between remittances and financial development. The authors used the share of bank deposits and the ratio of bank credit to the private sector expressed as a percentage of GDP as the proxies for financial development. By suing the data of 109 countries for the period 1975-2007, the study revealed a positive association between remittances and financial development.

Giuliano & Ruiz-Arranz (2009) examined how host country's financial sector development plays a crucial role to gain benefits from remittances. Using a dataset of 100 developing countries for the period 1975-2002, the study found that remittances boost growth in countries with less developed financial systems by providing an alternative way to finance investment. The study also concluded that remittances help alleviate credit constraints contributing to improve the allocation of capital and to boost economic growth.

Fromentin (2017) analyzed the relationship between remittances and financial development in emerging and developing countries during 1974-2014. The study found that there exists a significant positive relationship between remittances and financial development in the long run, in the short run, there exists a positive relationship between them except for low-income countries. Therefore, the study suggested that the policymakers should formulate and implement such policies so that the migrant workers would choose remit money through formal channels.

Bhattacharya et al. (2018) analyzed the association between remittances and financial development in developing and developed countries. The study revealed that there was a significant association between remittance inflows and financial development in the long run. They also suggested that the policymakers should lower the transaction costs of remittances to encourage more inflows of remittances through formal financial channels.

Kim (2021) studied the impact of remittances and institutional quality on financial development in developing countries. Using data of 46 countries for the period of 1996-2016, the author revealed that both institutional quality and remittances had a positive relationship with financial development. The study also revealed that financial openness and trade openness are shown to be critical for financial development.

## 3. Data and Methodology

In our study, we used Financial Development as a dependent variable and remittances as the variable of interest i.e., the main control variable. To explore the correlation between remittances and financial development, I estimated the following equation:

$$FD_{i,t} = \beta_0 + \beta_1 Rem_{i,t} + \beta_2 x_{i,t} + \varepsilon_{i,t}$$  (1)

Where, i indicates country and t indicates year. Xi,t represents the set of control variables which can affect financial development. Xi,t includes inflation which is an important indicator of overall macroeconomic situation of an economy (Barro, 1995; Biswas, 2023b). Foreign Direct Investment (FDI), Openness to trade are other control variables of my study as these variables reflect the performance of external sector of any economy (Borensztein et al., 1998; Li & Liu, 2005). Openness to trade has been calculated as import plus export divided by GDP (Yanikkaya, 2003). I used consumer price index (CPI) as the proxy for inflation and I have used broad money (M2) as a percentage of GDP as the proxy for financial development, which is also the dependent variable in my econometric analysis. The above-mentioned variables during 1996-2019 have been derived from World Bank Indicator (WDI) of the World Bank. Lastly, et represents an error term.

## 4. Empirial Results

*4.1 Trends of Remittances*

The Fig.1 shows the trends of remittances in these countries.

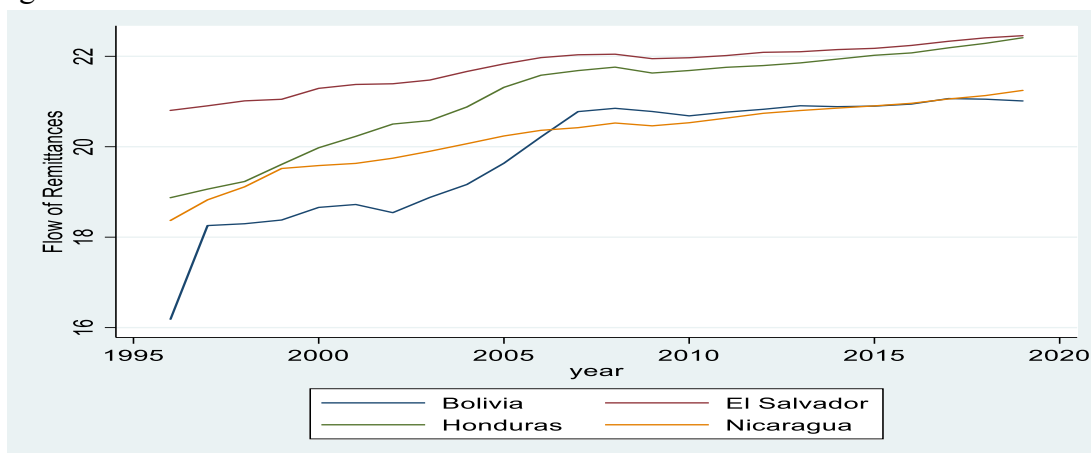

Fig. 1: Trends of Remittances Flow

The above Fig.1 shows the trend of remittance inflow in these selected countries. As we see from the figure that remittance flows in all these countries increases over the time. Among these countries, El Salvador has the highest remittance flows over the time, whereas Bolivia is the country with having lowest remittance inflows over the time except few years in the middle of the period.

## 4.2 Descriptive Statistics

The below Table 1 presents the descriptive statistics.

Table 1: Descriptive statistics

| Variable | Observation | Mean | Std. Dev. | Min | Max |
|---|---|---|---|---|---|
| L_bm | 96 | 3.89 | 0.32 | 3.02 | 4.56 |
| L_rem | 96 | 20.75 | 1.22 | 16.18 | 22.46 |
| L_inf | 96 | 4.46 | 0.39 | 3.35 | 5.09 |
| L_fdi | 96 | 1.32 | 0.70 | -0.54 | 2.50 |
| openness | 96 | 0.84 | 0.25 | 0.44 | 1.37 |

From Table 1, we observe that the mean of broad money is 3.89, with minimum 3.02 and maximum 4.56. We also observe that there are also differences in logarithmic of inward remittances among these countries with standard deviation 1.22. The minimum of logarithmic of remittances is 16.18 and the maximum value is 22.46. While there is less differences in logarithmic of inflation and openness among these countries, with standard deviations for these are 0.39 and 0.25 respectively.

## 4.3 Correlation Matrix

Correlation matrix between the variables is shown in the below table:

Table 2: Correlation Matrix

| Variables | L_bm | L_rem | L_inf | L_fdi | openness |
|---|---|---|---|---|---|
| L_bm | 1.00 | | | | |
| L_rem | 0.36 | 1.00 | | | |
| L_inf | 0.45 | 0.67 | 1.00 | | |
| L_fdi | -0.25 | -0.21 | -0.007 | 1.00 | |
| openness | -0.16 | 0.36 | 0.08 | 0.27 | 1.00 |

As can be seen from the Table 2 that there is a positive correlation between remittances and broad money, i.e. financial development in these countries which is 0.36. Inflation also has positive association with broad money, and the value is 0.45, inflation and remittances are also positively correlated with broad money. Whereas foreign direct investment has negative correlation with broad money, remittances, and inflation. Finally, trade openness has negative correlation with broad money, but has positive correlation with remittances, inflation, and foreign direct investment.

## 4.4 Regression Results

In our study, we used a panel data set, and for regression analysis of panel data fixed effect and random effect estimation are two popular methods. Firstly, I estimated the fixed effect model of the equation (1), which is shown below Table 3:

Table 3: Fixed Effect Estimation

| Variables | Dependent Variable: l_bm |
|---|---|
| L_rem | 0.08*** |
| | (0.02) |
| L_inf | 0.017*** |
|  | (0.05) |
| L_fdi | 0.02 |
|  | (0.01) |
| openness | -0.21** |
|  | (0.08) |
| Observations | 89 |
| R-squared | 0.68 |

Note: *** = 1%, ** = 5%, * = 10% significance level

From the result of fixed effect estimation, we see that remittances had a positive effect on financial development. i.e. on broad money, which is statistically significant at 1% significance level. Therefore, 1 % increase in remittances would cause broad money to increase by 8%. Inflation is also positively related with financial development. Finally, the result says that foreign direct investment has positive effect on financial development as well, but this effect is not statistically significant and trade openness has a negative influence with financial development, which is also statistically significant.

In this section, I estimated the random effect model of equation (1), which is presented in the below Table 4:

Table 4: Random Effect Estimation

| Variables | Dependent Variable: l_bm |
|---|---|
| L_rem | 0.008 |
|  | (0.03) |
| L_inf | 0.33*** |
|  | (0.10) |
| L_fdi | -0.07* |
|  | (0.04) |
| openness | -0.29** |
|  | (0.13) |
| Observations | 89 |
| R-squared | 0.52 |

Note: *** = 1%, ** = 5%, * = 10% significance level

From the result of random effect model, we see that remittances had a positive impact on financial development, but this effect is not statistically significant. Among other control variables, inflation had a positive relationship with financial development, which is statistically significant. Whereas foreign direct investment and trade openness had a negative relationship with financial development, and these effects are statistically significant.

*4.5 Hausman Test*

The results of fixed effect and random effect estimations are mixed. Hence, in this part, I conducted the Hausman test to check which model is appropriate. The result of the Hausman test is shown below:

Table 5: Hausman Test

| Test Summary | Chi-Sq statistics | Prob. |
|---|---|---|
| Cross-Section Random | 32.45 | 0.00 |

From the table we see that the p-value is 0.00, it means that we reject the null hypothesis that the random effect model is appropriate. Therefore, in my study the fixed effect model is appropriate.

## 5. Conclusion

Inward remittances have been considered as one the major drivers for any country, especially for the developing countries as remittances is an alternative source of investment. As developing countries have resource constraints, these countries do not have enough fund for investment activities. Therefore, along with foreign direct investment, remittances also play significant roles in accelerating economic activities in these countries, and Latin America countries are the major recipients of remittances in the world, hence in this paper, we analyzed the impact of remittances on financial development in 4 countries of Latin America. By using a panel data set from 1996-2019, the study found that remittances had a significant positive impact on financial development in these countries. Among other control variables, inflation and foreign direct investment had a positive impact, whereas trade openness created unfavorable condition on financial development.

The findings of the study have important implications. The policymakers of these countries should formulate and implement such policies so that the migrant workers will have the incentives to send money through the formal financial system, which will broaden the effect of remittances on home countries' economy.